# Angle-Resolved Photoemission of Solvated Electrons in Sodium-Doped Clusters

*Adam H. C. West, Bruce L. Yoder, David Luckhaus, Clara-Magdalena Saak, Maximilian Doppelbauer, and Ruth Signorell**

ETH Zürich, Laboratory of Physical Chemistry, Vladimir-Prelog-Weg 2, CH-8093, Zürich, Switzerland

* To whom correspondence should be addressed. E-mail: rsignorell@ethz.ch

**ABSTRACT**

Angle-resolved photoelectron spectroscopy of the unpaired electron in sodium-doped water, methanol, ammonia, and dimethyl ether clusters is presented. The experimental observations and the complementary calculations are consistent with surface electrons for the cluster size range studied. Evidence against internally solvated electrons is provided by the photoelectron angular distribution. The trends in the ionization energies seem mainly determined by the degree of hydrogen bonding in the solvent and the solvation of the ion core. The onset ionization energies of water and methanol clusters do not level off at small cluster sizes, but decrease slightly with increasing cluster size.

Solvated electrons have been studied extensively with various experimental and theoretical approaches, motivated by their relevance to liquid phase and radiation chemistry, biology, and potentially atmospheric processes.[1-12] Experimental cluster spectroscopy has contributed substantially to the characterization of the dynamic and structural properties of the solvated electron in different solvents including water, ammonia, and alcohols. Two different types of clusters have been investigated: anion clusters (refs.[13-25] and references therein), i. e. bare clusters with an excess electron, and neutral clusters doped with a single sodium (Na) atom (refs.[26-34] and references therein), in which the electron originates from the unpaired electron of the Na atom. Photodetachment of anion clusters revealed different types of isomers with two major structural features. Either the electron is located in the surface region or the electron is solvated inside the cluster.[14] Surface and internally solvated electrons were also found in liquid micro-jet photoemission studies,[35-42] albeit the surface electron seems to be only a transient short lived (sub-ps to ps) species in the micro-jet.[35, 38] These two types of isomers could not be identified in the photoionization studies of Na-doped clusters. It is still debated at which cluster sizes the internally solvated electron emerges in the different types of clusters. From the experimental data available, it also remains unclear whether extrapolation of data for large clusters to infinite size is representative for bulk behavior. Obvious issues arise from differences in the temperature and phase of clusters and bulk (solid vs. liquid). Less obvious is the potential influence of counter ions that are present, such as the Na cation in Na-doped clusters.

This Letter presents results from angle-resolved photoelectron spectroscopy (PES) of Na-doped water ($Na(H_2O)_n$), methanol ($Na(CH_3OH)_n$), ammonia ($Na(NH_3)_n$), and dimethyl ether ($Na(CH_3OCH_3)_n$) clusters as a function of solvent cluster size. The present investigation complements previous results from photoionization studies of Na-doped clusters, which all report "ion appearance energies" (AE) as a function of cluster size (ref.[26] and references therein). Angle-resolved PES of Na-doped clusters have not been recorded before and the few existing PES studies did not report quantitative data or focused on different subjects.[43, 44] Compared with photoionization spectroscopy where ions are detected, PES provides direct information on the ionization energies (IE). Furthermore, angle-resolved PES allows us to exploit the additional information that is contained in the photoelectron angular distribution (PAD). A very recent angle-resolved PES study of solvated electrons in a liquid water micro-jet revealed distinct differences in the PAD depending on the location of the electron relative to the liquid jet surface.[35] With the present experimental approach, we address open questions concerning electron solvation in Na-doped clusters with regard to the location of the electron, the influence of the Na counter ion, and the role of the degree of hydrogen-bonding in the solvent cluster. Unpaired electrons that reside near the surface are referred to as "surface electrons" and electrons that reside inside the cluster are referred to as "internally solvated electrons" throughout this Letter.

Details and schemes of the velocity map imaging (VMI) photoelectron spectrometer have been described previously.[45-47] Solvent clusters of different size (n = 1 to a few hundred; with n indicating the number of monomers) were generated by supersonic expansions of sample gases (neat or seeded in He/Ar/$N_2$) through a small nozzle into vacuum. After passing the skimmer, the solvent clusters picked up a single Na-atom in the Na pick-up cell (heated Na-oven) and entered the ionization chamber. Photons of 4.66 eV energy from a Nd:YAG laser (repetition rate 20 Hz) were used to ionize the unpaired electron of the Na in the neutral Na-doped clusters. A VMI setup was used to image the photoelectrons onto microchannel plates which were coupled to a phosphor screen with CCD camera.[48-50] The ionization energy (IE) and the photoelectron angular distribution (PAD) were determined from reconstructed images. The IE is the difference between the photon energy and the measured electron kinetic energy. Reconstructions were performed with BASEX[51] and MEVIR.[52] Two types of IEs are defined for each cluster size. The energy at the onset ($IE_{on}$) of the corresponding band and the energy at the band maximum ($IE_{max}$) (see Supporting Information). The PAD

$$\frac{d\sigma}{d\Omega} = \frac{\sigma_{tot}}{4\pi}\left[1 + \frac{\beta}{2}\left(3\cos^2\theta - 1\right)\right] \qquad (1)$$

is characterized by the anisotropy parameter $\beta$,[53] where $\frac{d\sigma}{d\Omega}$ and $\sigma_{tot}$ are the differential and the total photoionization cross section, respectively, and $\theta$ is the angle between the photoelectron velocity vector and the polarization axis of the incident light. To complement our experimental results, we have performed density functional (DFT) calculations for structures, vertical IE (VIE), and adiabatic IE (AIE) for clusters of up to 40 molecules. $\beta$ parameters were calculated on the basis of the DFT calculations following the model suggested in refs.[54, 55] (see Supporting Information for details and refs.[35, 56, 57]).

Figure 1 presents typical photoelectron images and photoelectron spectra for the examples of $Na(CH_3OCH_3)_n$ clusters. The images and PES clearly show that the PAD remains anisotropic for all cluster sizes and that the IEs decrease with increasing cluster size, respectively. Similar images and PES are found for the three other solvents $H_2O$, $CH_3OH$, and $NH_3$. Figure 1 shows data for clusters with n = 1 to 6 solvent molecules and for a cluster size distribution with an average number of solvent molecules of $n_{av} = 37$. Size-resolved spectra are obtained up to n ~ 6. For larger solvent clusters, the spectra are averages over distributions of different cluster sizes. In contrast to anion clusters, size selection before the photoemission experiment is not possible for the neutral Na-doped clusters. However, it is possible to determine an accurate size distribution for each photoelectron spectrum using time-of-flight mass spectrometry as described in refs.[45, 58] The size assignment is performed on the basis of these mass spectra (see Figure S1 and Table S1 in the Supporting Information). Even without size selection the present study demonstrates that clear size-dependent trends can be probed by PES also for neutral clusters. Figures 2a and 2b show a summary of the evolution of $IE_{on}$ and $IE_{max}$, and of the anisotropy parameter $\beta$, respectively, as a function of cluster size for the four different solvents. (The corresponding data are provided in Table S1 in the Supporting Information). All IEs decrease with increasing solvent cluster size, although to a varying degree. Overall, the $IE_{max}$ reproduce the trends (not the absolute values) observed by Buck, Hertel and coworkers for AEs of $H_2O$, $CH_3OH$, and $NH_3$, namely that plateaus are reached for $H_2O$ (at n = 4) and $CH_3OH$ (at n = 6) above a certain cluster size, while the values for $NH_3$ continue to decrease.[26] The $IE_{on}$ of $H_2O$ and $CH_3OH$ clusters, by contrast, seem not to reach plateau values. They slightly decrease further with increasing cluster size. This is a new behavior that has not been observed before for the AE. It is, however, similar to the trends predicted from calculations for the adiabatic ionization potentials (AIE) of $Na(H_2O)_n$ clusters.[59] It is important to note here that the measured $IE_{max}$ and $IE_{on}$ need not necessarily represent adiabatic and vertical ionization energies, respectively. The fourth solvent, $CH_3OCH_3$, has not been studied previously. It is an example of a solvent where hydrogen bonding is negligible and which does not dissolve Na in its liquid bulk (see comments in the Supporting Information). The comparison of the IEs of the four solvents shows a clear trend: the weaker the hydrogen bonding of the solvent the more pronounced the decrease of the IE as a function of cluster size. For the $\beta$ parameters in Figure 2b, there is no obvious correlation with the strength of the hydrogen bonding. The only trend is that all $\beta$s are larger than zero, even for the larger clusters.

For all solvents and cluster sizes studied here (up to n = 40), the DFT calculations predict that the unpaired electron is located at the surface as illustrated for dimethyl ether and ammonia in Figure 3a. In the following, we focus on these two substances because of their opposite bulk behavior; i. e. bulk ammonia dissolves Na whilst dimethyl ether does not. The comparison of experimental and calculated IEs in Figure 3b demonstrates that the evolution of the IEs can be explained by structures with a surface electron and does not require the formation of an internally solvated electron. Furthermore, dimethyl ether, which does not dissolve Na in the bulk and has almost no hydrogen bonding, shows the largest drop in IE (Figure 2a). An interpretation of the decrease in the IEs in terms of the emergence of the internally solvated electron is clearly too simplistic. The calculations also reproduce the experimental trend observed as a function of the strength of the hydrogen-bond network (Figure 2a). This implies that the main contribution to the decrease of the IE arises from the solvation of the charged Na core. According to our calculations the simple picture of a Na atom losing its electron upon solvation by a molecular cluster is not correct. In the cluster size range considered, the HOMO reveals an unpaired surface electron that largely retains the character of a sodium valence electron. So instead of an independently solvated electron and ion we have the solvation of a polarized sodium atom with a distorted valence electron residing on the surface and a (partially) charged Na-core ($Na^{\delta+}$) inside the cluster. For strong hydrogen bonding as in $H_2O$, the solvation of the $Na^{\delta+}$ disturbs the hydrogen-bond network. This costs energy and results in a less pronounced decrease of the IE with increasing solvent cluster size. In $CH_3OCH_3$, by contrast, no hydrogen-bond network needs to be disturbed. The $Na^{\delta+}$ solvation is optimal in this case, which results in a very strong decrease of the IE. $CH_3OH$, and $NH_3$ represent intermediate cases. A better solvated $Na^{\delta+}$ corresponds to a more weakly bound electron. The actual location of the electron – inside or at the surface – would be of minor importance for the evolution of the IE. This effect is also reflected in the reorganization energy upon ionization, i.e. the difference between VIE and AIE, which is largest for the strongly H-bonded systems and virtually zero for $CH_3OCH_3$, with $CH_3OH$ and $NH_3$ in between. The situation in Na-doped clusters is thus fundamentally different from anion clusters, which has not always been recognized. Electron-solvent interaction is crucial for the latter, while the balance of the additional interactions of the electron and of the solvent with the ionic core determines the behavior of Na-doped clusters.

The anisotropy in the PAD helps us to further elucidate the nature of the unpaired electron in the Na-doped clusters. The comparison of the calculated $\beta$ parameters and the experimental ones in Figure 3c is also consistent with a surface electron. The trends calculated for surface electron structures are qualitatively similar to the experimental observations. The major differences are the higher absolute values for $\beta$ in the calculations. Since Na does not dissolve in bulk dimethyl ether and since the calculations for this solvent only support surface electrons, the evolution of the experimental $\beta$ parameters for $Na(CH_3OCH_3)_n$ clearly represents the evolution of the surface electron with size. The jump in $\beta$ observed for $Na(CH_3OCH_3)_6$ is the result of the near octahedral symmetry of this cluster. The calculations reveal that the HOMO of this cluster has pure s-character which results in a high $\beta$ value. For comparison, $Na(CH_3OCH_3)_5$ has an s-character of only about 55% (and 45% p-character) and consequently a lower anisotropy. According to the calculations, the observed drop of the $\beta$ parameter for $Na(NH_3)_n$ for the larger clusters results from a decrease in the s-character and an increase in the p-character of the surface electron, and not from the formation of an internally solvated electron. The situation for the three biggest $Na(NH_3)_n$ clusters in Figure 3c is unclear. We do not have calculations for these cluster sizes. The low $\beta$ for these clusters could simply correspond to a further decrease of the s-character of the surface electron with increasing size or it could indicate the emergence of the internally solvated electron.

Further strong evidence for surface electrons comes from the anisotropy of the larger $Na(H_2O)_n$ clusters with over a hundred solvent molecules (see data for $n_{max}$ = 320 in Table S1 in the Supporting Information). For these clusters, we measure a $\beta$ value of ~ $0.4 \pm 0.1$. This value is very close to the value of $\beta = 0.3 \pm 0.1$ that was recently determined by Suzuki and coworkers for Rydberg states of DABCO molecules segregated on the surface of a liquid water micro-jet.[35] To confirm the surface character of this Rydberg electron they have also calculated the anisotropy of a 3s Rydberg state of DABCO fully hydrated by 64 water molecules. The calculated photoemission distribution of this internally solvated system was isotropic ($\beta = 0$). The finite photoemission anisotropy observed experimentally was thus assigned to a surface structure. Suzuki and coworkers also provide experimental anisotropy parameters for a fully hydrated electron that still experiences electrostatic interaction with $DABCO^+$. Consistent with their calculations, they find an isotropic photoemission for these internally solvated electrons in the experiment. Electron scattering during the migration of an internally solvated electron through the solvent after photoexcitation is expected to be a major reason for the loss of anisotropy of internally solvated electrons. We would like to note here that almost complete randomization of the PAD at similar electron energies was recently also observed for $(NH_3)_n$ clusters with only a few tens of molecules per cluster.[45] Electron scattering could be a potential explanation here as well. The nature of the DABCO Rydberg state in water resembles that of our $Na(H_2O)_n$ clusters. Both are Rydberg type electrons interacting with a counter-cation. For this reason, we think that the finite $\beta$ value for larger Na-doped water clusters clearly hints at a surface electron, in agreement with the observations for the other solvents and with the calculations.

In conclusion, we report the first angle-resolved photoelectron studies of different types of Na-doped solvent clusters. These studies address outstanding questions concerning electron solvation. Experiments and calculations for $Na(H_2O)_n$, $Na(CH_3OH)_n$, $Na(NH_3)_n$, and $Na(CH_3OCH_3)_n$ are consistent with surface electrons for the cluster size range studied. The assumption of an internally solvated electron is not required to explain the experimental trends. The analysis of the PAD provides useful information on the location of the electron and provides evidence against an internally solvated electron. The strong decrease in the IEs with increasing solvent cluster size, by contrast, must not be taken as an indicator of the location of the electron for Na-doped clusters. The systematic dependence of the IEs on the type of solvent and its decrease seems to be mainly determined by the degree of hydrogen bonding and the solvation of the ion core. In contrast to the observed ion appearance energies,[26] the $IE_{on}$ do not level off at small cluster sizes for $Na(H_2O)_n$ and $Na(CH_3OH)_n$. We would like to stress that for finite cluster size the behavior of anion clusters differs qualitatively from that of Na-doped clusters as their energetics are free from the dominating influence of the counter-cations. Extrapolation of anion cluster data to infinite size thus seems a more reasonable approach to predict the properties of the solvated electron in the bulk. Because of the qualitative difference between anionic and neutral clusters there is also no direct connection between the types of isomers (surface versus internally solvated electron) present in anions and the behavior observed for Na-doped clusters.

---

**Acknowledgment:** We thank Guido Grassi for experimental assistance with the dimethyl ether and ammonia bulk measurements and David Stapfer and Markus Steger for technical support. Financial support was provided by the ETH Zürich and the Swiss National Science Foundation under project no. 200021_146368.

**Supporting Information (SI) Available:** The SI contains a description of the calculations, a description of determination of the cluster size, a table with ionization energies and anisotropy parameters, and a description of the bulk experiments. This material is available free of charge via the Internet at http://pubs.acs.org.

**LIST OF FIGURES**

**Figure 1.** (a) Raw and reconstructed photoelectron images of $Na(CH_3OCH_3)_n$ clusters with n = 1 to 6 solvent molecules (left) and $n_{av}$ = 37 solvent molecules (right). The arrow indicates the direction of light polarization. (b) Corresponding photoelectron spectra.

**Figure 2.** (a) Maximum ionization energies $IE_{max}$ (full symbols) and onset ionization energies $IE_{on}$ (open symbols) for the four different solvents as a function of solvent cluster size. (b) Anisotropy parameter $\beta$ for the four different solvents as a function of cluster size. n is the number of solvent molecules in a cluster. Note that n = $n_{av}$ for the larger clusters. See Table S1 in Supporting Information for data and errors.

**Figure 3.** (a) Isosurfaces for the HOMO of $Na(CH_3OCH_3)_{15}$ (left) and $Na(NH_3)_{30}$ (right) (b) Comparison between experimental ionization energies ($IE_{max}$ and $IE_{on}$) and calculated vertical (VIE) and adiabatic (AIE) ionization energies for dimethyl ether clusters (left) and ammonia clusters (right). (c) Comparison between experimental and calculated $\beta$ parameters for dimethyl ether clusters (left) and ammonia clusters (right).

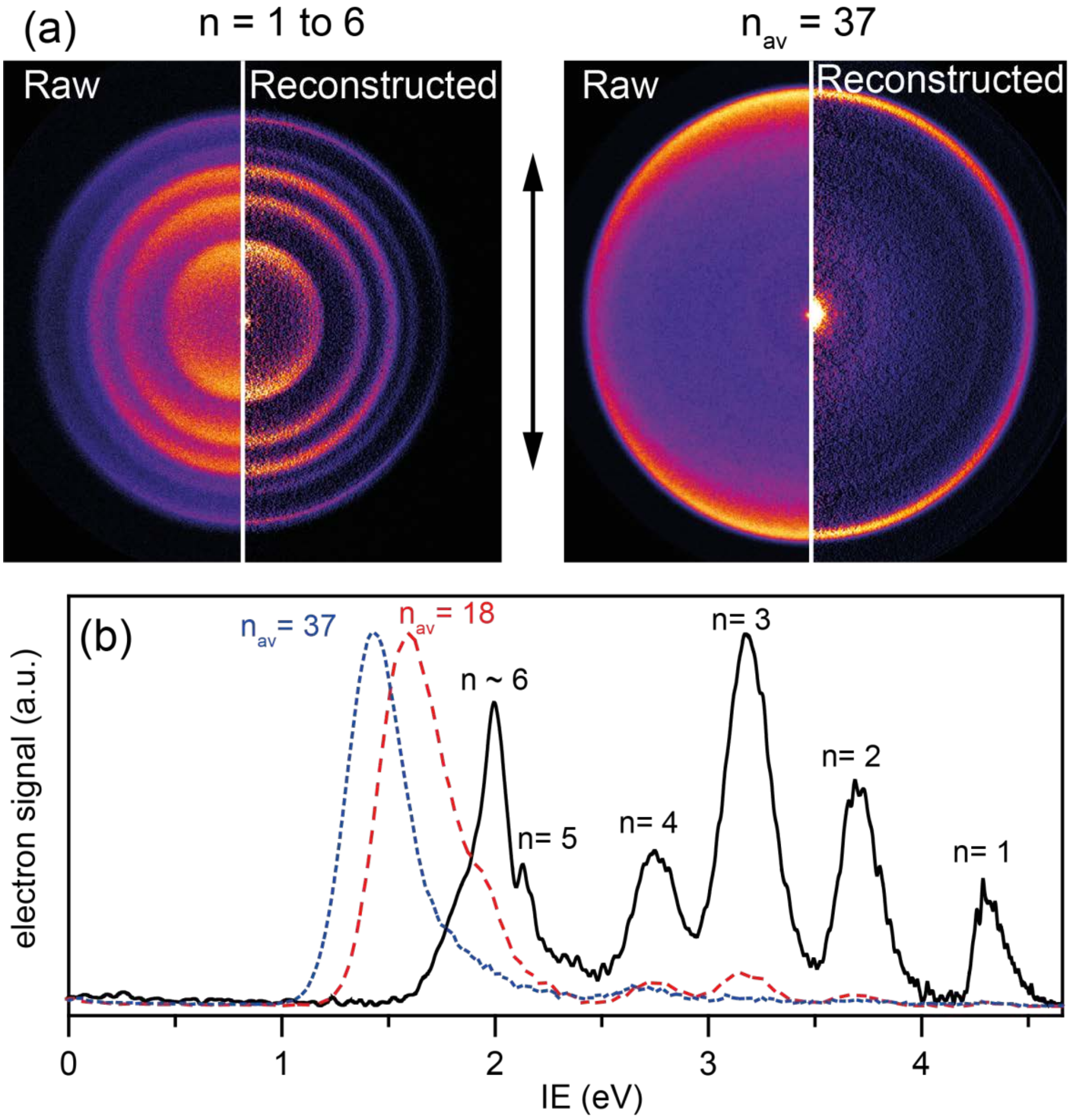

**Figure 1.** (a) Raw and reconstructed photoelectron images of $Na(CH_3OCH_3)_n$ clusters with n = 1 to 6 solvent molecules (left) and $n_{av}$ = 37 solvent molecules (right). The arrow indicates the direction of light polarization. (b) Corresponding photoelectron spectra.

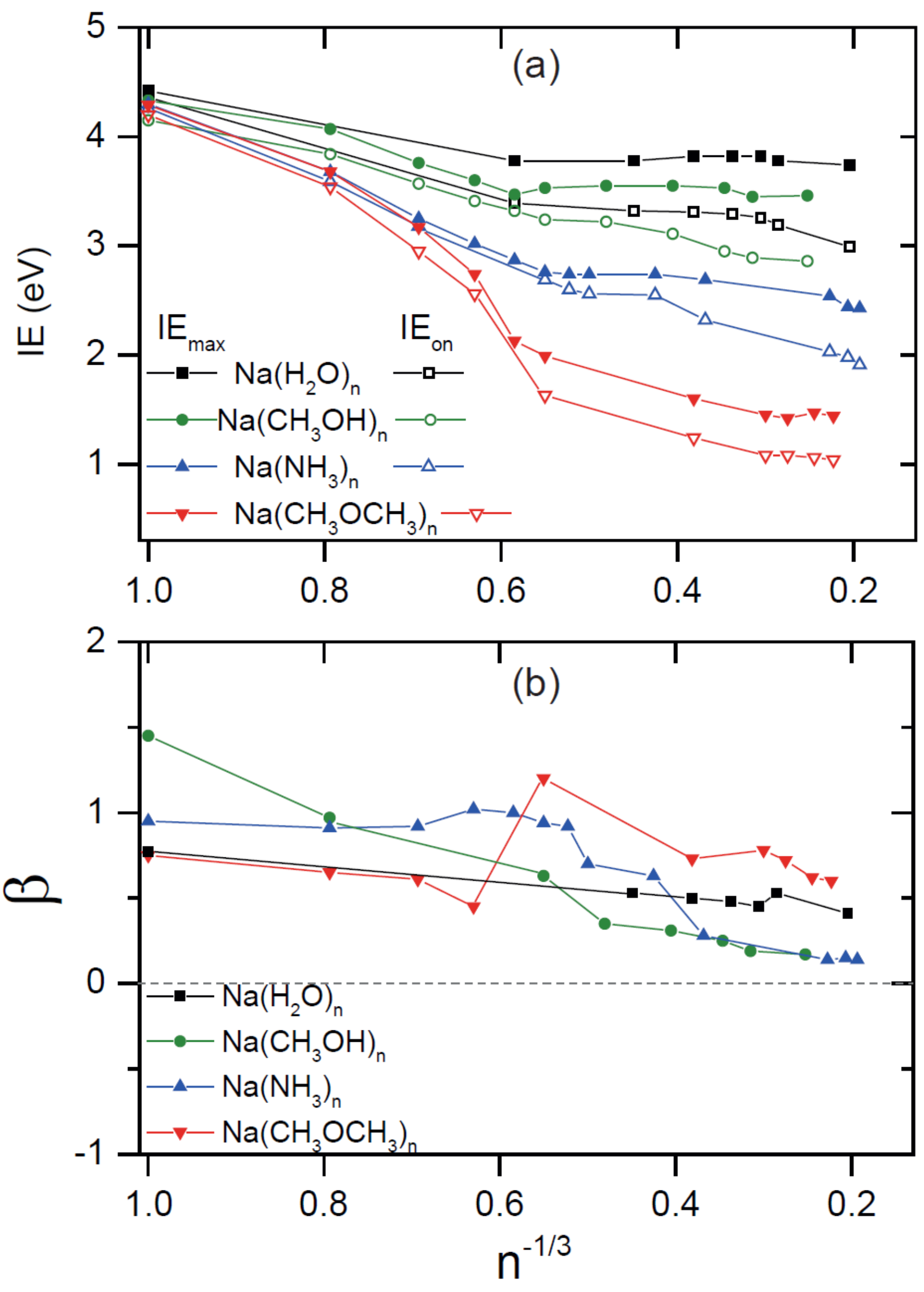

**Figure 2.** (a) Maximum ionization energies $IE_{max}$ (full symbols) and onset ionization energies $IE_{on}$ (open symbols) for the four different solvents as a function of solvent cluster size. (b) Anisotropy parameter $\beta$ for the four different solvents as a function of cluster size. n is the number of solvent molecules in a cluster. Note that n = $n_{av}$ for the larger clusters. See Table S1 in Supporting Information for data and errors.

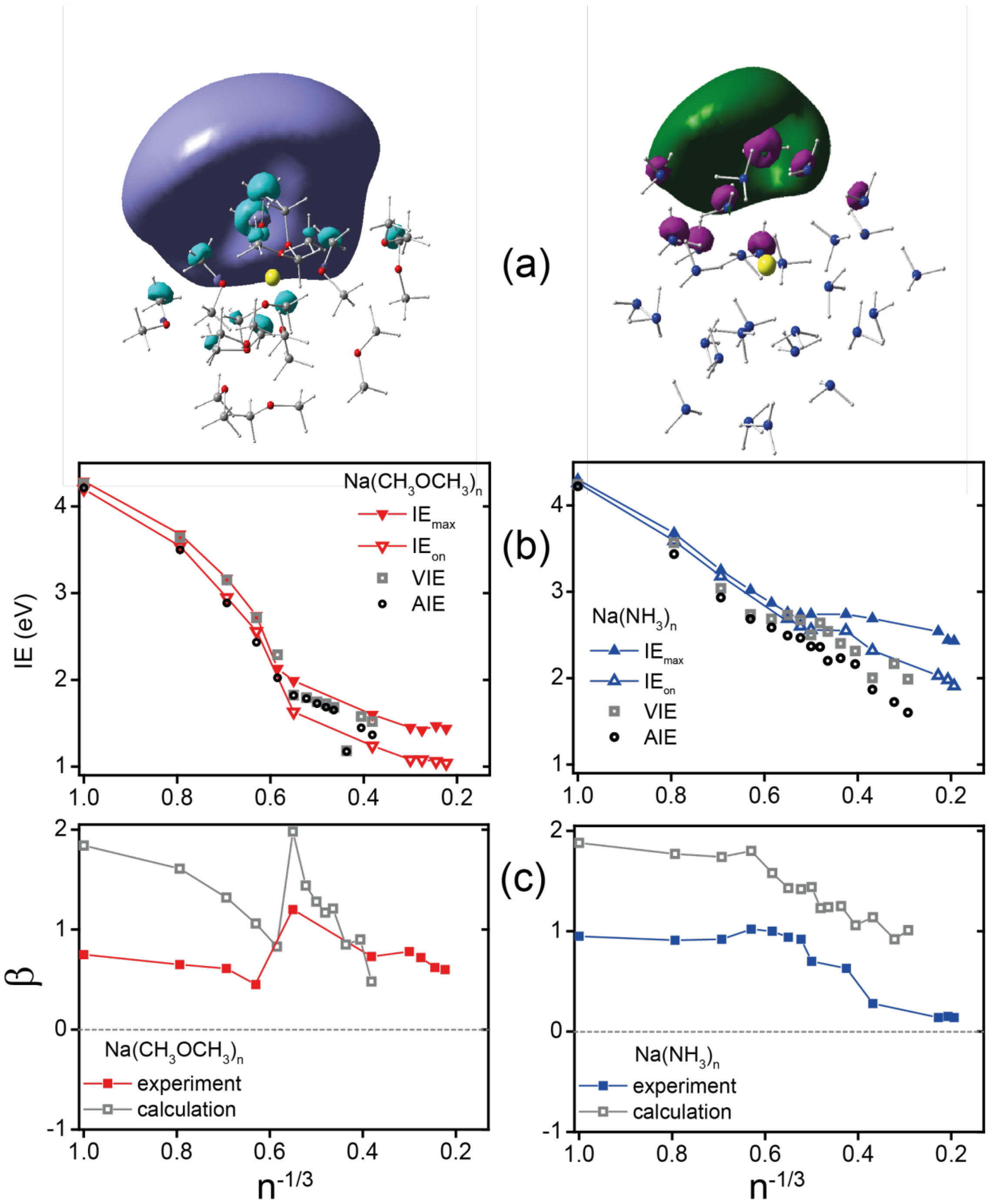

**Figure 3.** (a) Isosurfaces for the HOMO of $Na(CH_3OCH_3)_{15}$ (left) and $Na(NH_3)_{30}$ (right) (b) Comparison between experimental ionization energies ($IE_{max}$ and $IE_{on}$) and calculated vertical (VIE) and adiabatic (AIE) ionization energies for dimethyl ether clusters (left) and ammonia clusters (right). (c) Comparison between experimental and calculated $\beta$ parameters for dimethyl ether clusters (left) and ammonia clusters (right).

# Supporting Information for:

# Angle-Resolved Photoemission of Solvated Electrons in Sodium-Doped Clusters

*Adam H. C. West, Bruce L. Yoder, David Luckhaus, Clara-Magdalena Saak, Maximilian Doppelbauer, and Ruth Signorell**

ETH Zürich, Laboratory of Physical Chemistry, Vladimir-Prelog-Weg 2, CH-8093, Zürich, Switzerland

*rsignorell@ethz.ch

## 1. Description of calculations

Optimized structures of undoped, Na-doped neutral, and Na-doped ionic clusters were calculated with the Gaussian program package[1] for up to n = 40 solvent molecules using the dispersion corrected wB97XD density functional with a 6-31+G* basis set. To confirm that this approach yields reliable equilibrium geometries and an acceptable quantitative description of the balance between solvent-solvent, Na-solvent, and $Na^+$-solvent interaction the results were checked against higher levels of theory (2$^{nd}$ order Møller-Plesset perturbation theory with up to quadruple-$\zeta$ correlation consistent basis sets) for clusters up to n = 3.

Vertical (VIE) and adiabatic (AIE) ionization energies were determined by subtracting the energy of the neutral cluster from the energy of the ionic cluster with the same geometry and most stable relaxed geometry, respectively. For comparison of the calculated ionization energies with the experimental data, the lowest energy isomers of the neutral Na-doped clusters were chosen and the ionization energies referred to that of free Na. The corresponding data are shown in Figure 3b in the main text. Typical variations in the IEs of different isomers are 0.5-1.0 eV for strongly hydrogen bonded and 0.1-0.2 eV for weakly hydrogen bonded Na-doped clusters.

The $\beta$-parameters were calculated following the approach proposed by Melko and Castleman[2]:

$$\beta = \sum_{\ell} c_{\ell}^2 \beta_{\ell} \qquad \text{Eq. (S1)}$$

The sum extends over the angular momentum components $\ell$. $c_{\ell}^2$ is the normalized $\ell$-character of the highest occupied molecular orbital (HOMO) and is calculated as the sum over atomic natural orbital (ANO) contributions with the same $\ell$. The latter are obtained by expanding the HOMO in terms of ANOs[3], which are invariant to basis set transformations. The $\beta_{\ell}$ were determined from the Cooper–Zare formula[4] neglecting the phase shift between outgoing partial waves,

$$\beta_\ell = \frac{\ell(\ell-1)(1-R)^2 + (\ell+1)(\ell+2)R^2 - 6\ell(\ell+1)(1-R)R}{(2\ell+1)\left[\ell(1-R)^2 + (\ell+1)R^2\right]} \qquad \text{Eq. (S2)}$$

$R$ is the relative radial dipole matrix element of the ($\ell$+1) partial wave with $R$=0.5 for the "maximum interference" case[2], for which the results are shown in Figure 3c in the main text. Note that Eq. (S1) retains the form of the exact expression (eq. (21) in ref.[5]). The only difference is that Eq. (S2) omits an $\ell$ -dependent scaling of the $\beta_\ell$. It depends on the absolute values of the radial dipole matrix elements, which are unknown. With the assumption of a central potential our model represents a drastic simplification, so that we do not expect it to reproduce absolute $\beta$-parameters. Considering the limiting cases of $R$=0 (pure p→s), $R$=1 (pure p→d), and $R$=0.5 ("maximum interference")[2], however, we find the same qualitative behaviour of $\beta$ as a function of cluster size. This robustness with respect to the unknown values of the radial dipole matrix elements gives us confidence in the predicted qualitative cluster size dependence. Figure 3c compares the "maximum interference" case – again for the most stable Na-doped cluster isomer – with our experimental data. The size dependence of $\beta$ is robust with respect to the choice of parameters because the HOMO is dominated by Na-contributions: The outgoing electron experiences a largely atomic environment, which is modulated by the solvent cluster, but remains qualitatively similar for all cluster sizes. We tested this explanation by restricting the summation in Eq. (S1) to Na ANO contributions. The resulting $\beta$ values show the same qualitative behaviour with respect to cluster size as the full calculation. By the same argument, the size dependence of $\beta$ is not much affected by the phase shift of the outgoing partial waves. The neglect of the phase shift is usually justified for photodetachment processes by the weak interaction between the outgoing electron and the neutral core. This does obviously not hold for photoionization where the core is charged. But again, in Na-doped clusters the ionization process in essence remains that of a Na atom for all cluster sizes. While appreciable phase shifts would affect the absolute $\beta$-values, they would do so to a similar extent for all cluster sizes with correspondingly little influence on the size dependence of $\beta$. According to the above arguments, the size-dependence of $\beta$ is dominated by the $\ell$-character of the HOMO.

## 2. Determination of cluster size and table of ionization energies and anisotropy parameters

**Determination of cluster size:**

The cluster size distributions were determined from the mass spectra (see main text and references therein). As an example, Fig. S1 shows two typical mass spectra for Na-doped dimethyl ether clusters. The ionization energies (IE) of the small clusters (typically $n < 6$ where n is the number of solvent monomers in the cluster) differ substantially. For these small clusters, we have recorded photoelectron spectra (PES) for a range of different cluster size distributions which differed in the relative contributions of clusters of size n. This procedure allowed us to assign a certain band in the photoelectron spectrum to a specific cluster size n. For the larger clusters, this is no longer possible. Here, we determined the average (count mean) number of solvent molecules per cluster $n_{av}$ from the mass spectrum and assigned this average size to the corresponding PES (see Table S1 and Figure 1 in the main text). The maximum number of solvent molecules per cluster $n_{max}$ is also listed in Table S1 to provide an impression of the widths of the cluster size distributions. $n_{max}$ was determined as the largest cluster size that has a signal two times higher than the noise level.

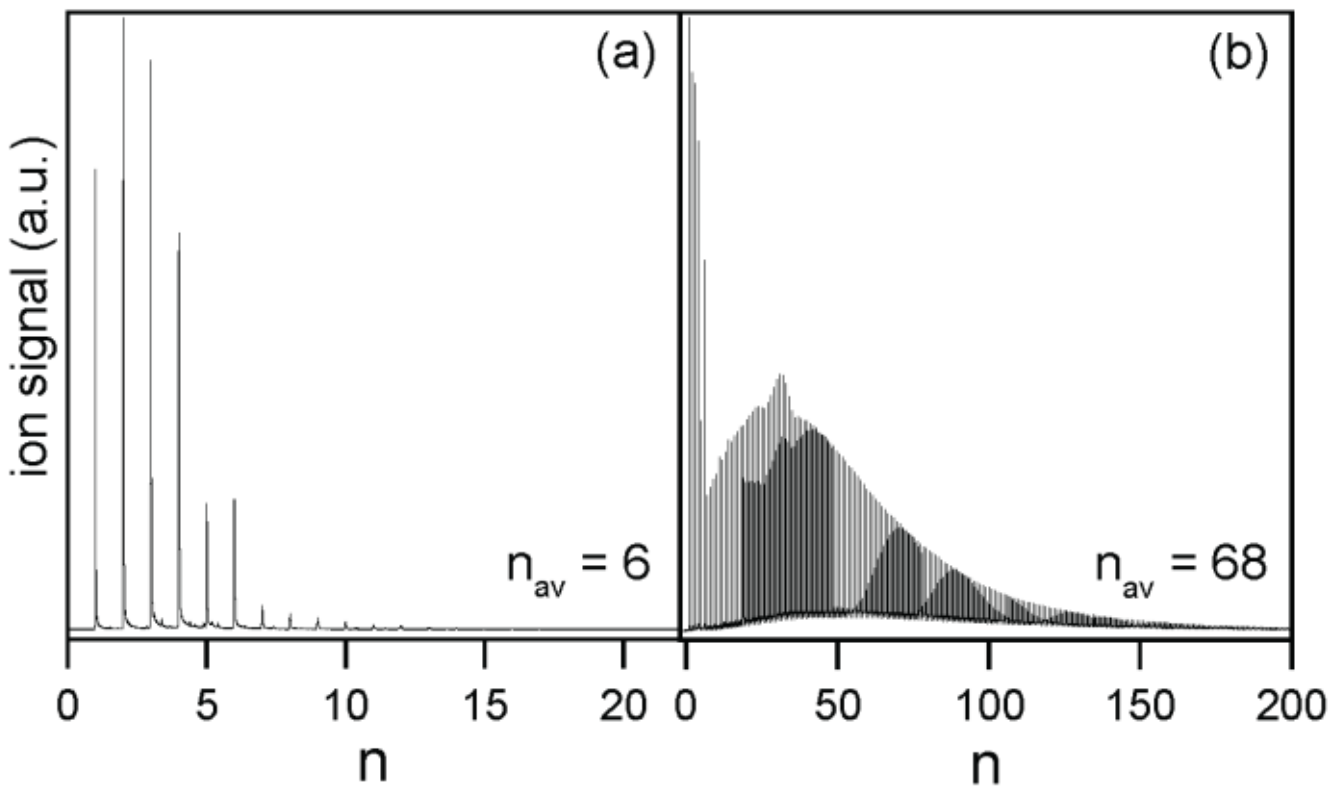

**Figure S1: Representative mass spectra of $Na(CH_3OCH_3)_n$ clusters.**

## Table of ionization energies and anisotropy parameters:

Table S1 contains the values for the IEs and β-parameters displayed in Figure 2 in the main text. Two different reconstruction methods, BASEX[6] and MEVIR[7], were used. The two methods provided similar results except for images with poor signal-to-noise, where MEVIR has been shown to yield more accurate speed distributions than other reconstruction methods[7], and indeed produced both speed distributions and anisotropy parameter distributions with less noise than BASEX. All β-parameters shown here are determined with MEVIR.

$IE_{max}$ is the IE at a given band maximum in the photoelectron spectrum. $IE_{on}$ is the ionization energy on the low energy side of the band where the electron signal rises twice above the noise level. We used Gaussian fits to determine the IEs from the experimental PES. The uncertainties in $IE_{max}$ and $IE_{on}$ indicated in Table S1 arise from uncertainties in the energy calibration, center-spot determination, different reconstruction methods, roundness of the image, and errors in the Gaussian fits.

The β-parameter for a given band in the PES was determined as the average over 2.5 pixels on either side of the band maximum. The uncertainties indicated in Table S1 reflect uncertainties associated with different reconstruction methods, the roundness of the image, as well as the fitting procedure.

Note that the calculations in Figure 3 are for a single cluster size $n_{av}$ and that explicit averaging over the cluster size distribution does not lead to any significant difference.

| $Na(H_2O)_n$ | | | | |
|---|---|---|---|---|
| $n_{max}$ | n (either n or $n_{av}$) | $IE_{max}$ (eV) | $IE_{on}$ (eV) | β |
| 1 | 1 | 4.42 ± 0.10 | 4.36 ± 0.10 | 0.77 ± 0.32 |
| 10 | 5 | 3.78 ± 0.10 | 3.39 ± 0.10 | -- |
| 31 | 11 | 3.78 ± 0.10 | 3.32 ± 0.10 | 0.53 ± 0.16 |
| 68 | 18 | 3.82 ± 0.10 | 3.31 ± 0.10 | 0.50 ± 0.10 |
| 75 | 26 | 3.82 ± 0.10 | 3.29 ± 0.10 | 0.48 ± 0.14 |
| 86 | 35 | 3.82 ± 0.10 | 3.26 ± 0.10 | 0.45 ± 0.10 |
| 102 | 43 | 3.78 ± 0.10 | 3.19 ± 0.10 | 0.53 ± 0.16 |
| 320 | 117 | 3.74 ± 0.10 | 2.99 ± 0.10 | 0.41 ± 0.10 |
| $Na(CH_3OH)_n$ | | | | |
| $n_{max}$ | n (either n or $n_{av}$) | $IE_{max}$ (eV) | $IE_{on}$ (eV) | β |
| 1 | 1 | 4.33 ± 0.10 | 4.15 ± 0.10 | 1.45 ± 0.18 |
| 2 | 2 | 4.07 ± 0.10 | 3.84 ± 0.10 | 0.97 ± 0.28 |
| 3 | 3 | 3.76 ± 0.10 | 3.57 ± 0.10 | -- |
| 4 | 4 | 3.60 ± 0.10 | 3.41 ± 0.10 | -- |
| 5 | 5 | 3.47 ± 0.10 | 3.32 ± 0.10 | -- |
| 11 | 6 | 3.53 ± 0.10 | 3.24 ± 0.10 | 0.63 ± 0.24 |
| 25 | 9 | 3.55 ± 0.10 | 3.22 ± 0.10 | 0.35 ± 0.10 |
| 43 | 15 | 3.55 ± 0.10 | 3.11 ± 0.10 | 0.31 ± 0.10 |
| 69 | 24 | 3.53 ± 0.10 | 2.95 ± 0.10 | 0.25 ± 0.10 |
| 109 | 32 | 3.45 ± 0.10 | 2.89 ± 0.10 | 0.19 ± 0.10 |
| 192 | 62 | 3.46 ± 0.10 | 2.86 ± 0.10 | 0.17 ± 0.10 |
| $Na(NH_3)_n$ | | | | |
| $n_{max}$ | n (either n or $n_{av}$) | $IE_{max}$ (eV) | $IE_{on}$ (eV) | β |
| 1 | 1 | 4.30 ± 0.10 | 4.26 ± 0.10 | 0.95 ± 0.24 |
| 2 | 2 | 3.68 ± 0.10 | 3.59 ± 0.10 | 0.91 ± 0.10 |
| 3 | 3 | 3.25 ± 0.10 | 3.18 ± 0.10 | 0.92 ± 0.16 |
| 4 | 4 | 3.02 ± 0.10 | -- | 1.02 ± 0.12 |
| 5 | 5 | 2.87 ± 0.10 | -- | 1.00 ± 0.10 |
| 9 | 6 | 2.76 ± 0.10 | 2.69 ± 0.10 | 0.94 ± 0.32 |
| 18 | 7 | 2.74 ± 0.10 | 2.60 ± 0.10 | 0.92 ± 0.24 |
| 22 | 8 | 2.74 ± 0.10 | 2.56 ± 0.11 | 0.70 ± 0.10 |
| 40 | 13 | 2.74 ± 0.10 | 2.55 ± 0.11 | 0.63 ± 0.18 |
| 90 | 20 | 2.69 ± 0.10 | 2.32 ± 0.12 | 0.28 ± 0.10 |
| 170 | 85 | 2.54 ± 0.11 | 2.03 ± 0.13 | 0.14 ± 0.10 |
| 280 | 113 | 2.44 ± 0.11 | 1.98 ± 0.13 | 0.15 ± 0.10 |
| 400 | 138 | 2.43 ± 0.11 | 1.91 ± 0.14 | 0.14 ± 0.10 |
| $Na(CH_3OCH_3)_n$ | | | | |
| $n_{max}$ | n (either n or $n_{av}$) | $IE_{max}$ (eV) | $IE_{on}$ (eV) | β |
| 1 | 1 | 4.29 ± 0.10 | 4.20 ± 0.10 | 0.75 ± 0.20 |
| 2 | 2 | 3.68 ± 0.10 | 3.54 ± 0.10 | 0.65 ± 0.10 |
| 3 | 3 | 3.17 ± 0.10 | 2.95 ± 0.10 | 0.61 ± 0.10 |
| 4 | 4 | 2.74 ± 0.10 | 2.56 ± 0.11 | 0.45 ± 0.12 |
| 5 | 5 | 2.13 ± 0.13 | -- | -- |
| 12 | 6 | 1.99 ± 0.13 | 1.63 ± 0.15 | 1.20 ± 0.16 |
| 45 | 18 | 1.60 ± 0.15 | 1.24 ± 0.17 | 0.73 ± 0.16 |
| 65 | 37 | 1.45 ± 0.16 | 1.08 ± 0.18 | 0.78 ± 0.30 |
| 155 | 48 | 1.42 ± 0.16 | 1.08 ± 0.18 | 0.72 ± 0.30 |
| 270 | 68 | 1.47 ± 0.16 | 1.06 ± 0.18 | 0.62 ± 0.30 |
| 300 | 90 | 1.44 ± 0.16 | 1.04 ± 0.18 | 0.60 ± 0.10 |

**Table S1: Experimental data for Na-doped water, methanol, ammonia and dimethyl ether clusters. See text in this section for the error bars and text in section 2 for the definitions of n, $n_{av}$, and $n_{max}$.**

## 3. Experiments with sodium in liquid dimethyl ether or ammonia bulk

Remarks in the literature hint that Na cannot be dissolved in simple liquid ethers[8-11], such as dimethyl ether (note that in the references DME does not always refer to dimethyl ether). In order to confirm this we have carried out two experiments. In the first one, a piece of pure sodium (Na), ~ 2g, was placed into a round-bottomed flask containing ~ 100 mL liquid ammonia ($NH_3$) under a nitrogen atmosphere maintained at ambient pressure and cooled to -78°C in an ethanol/dry ice cooling bath. The Na dissolved immediately, producing an intense dark blue color associated with the formation of solvated electrons, which gradually grew darker as the Na continued to dissolve. In the second experiment, ~ 100 mL liquid dimethyl ether at -78°C was used instead of ammonia and the same procedure was carried out. Na did not dissolve in liquid dimethyl ether. The temperature was then slowly risen to the boiling point of DME (-24°C) over more than an hour. Again, dissolution of Na in liquid dimethyl ether was not observed. From the hints in the literature and from our experiments we thus conclude that liquid bulk dimethyl ether does not dissolve Na in any substantial amount.